\documentclass[journal]{IEEEtran}

\usepackage{amsmath}
\usepackage{graphicx}
\usepackage{subfigure}
\usepackage{multicol} 
\usepackage{epstopdf}
\usepackage{booktabs}  
\usepackage{threeparttable} 
\usepackage{multirow}
\usepackage{amssymb}
\usepackage{color}

\ifCLASSINFOpdf
\else
\fi
\begin{document}
%
\title{ Revealing Fine Structures of the Retinal Receptive Field by Deep Learning Networks}
%
%
%

\author{
Qi Yan, Yajing Zheng, Shanshan Jia, Yichen Zhang, 
Zhaofei Yu, 
Feng Chen, 
Yonghong~Tian, 
Tiejun~Huang, 
Jian~K.~Liu
	
\thanks{\textit{Qi Yan and Yajing Zheng contributed equally to this paper. Corresponding authors: Zhaofei Yu and Jian K. Liu.} }	
\thanks{Q. Yan is with the Department of Automation, Tsinghua University, Beijing, China. e-mail: (q-yan15@mails.tsinghua.edu.cn).}
\thanks{Y. Zheng, S. Jia, Y. Zhang, Z. Yu, Y. Tian, T. Huang are with the National Engineering Laboratory for Video Technology, School of Electronics Engineering and Computer Science, Peking University, Beijing, China. e-mail: (zyj061@pku.edu.cn, jssll00@126.com, zhangyichen2622@outlook.com, yuzf12@pku.edu.cn, yhtian@pku.edu.cn, tjhuang@pku.edu.cn).} 
\thanks{F. Chen is with the Department of Automation, Beijing Innovation Center for Future Chip, LSBDPA Beijing Key Laboratory, Tsinghua University, Beijing, China.. e-mail: (chenfeng@tsinghua.edu.cn).}
\thanks{J. K. Liu is with the Centre for Systems Neuroscience, Department of Neuroscience, Psychology and Behaviour, University of Leicester, Leicester, UK. e-mail: (jian.liu@leicester.ac.uk).}

}

\maketitle

\begin{abstract}
Deep convolutional neural networks (CNNs) have demonstrated impressive performance on many visual tasks. Recently, they became useful models for the visual system in neuroscience. However, it is still not clear what are learned by CNNs in terms of neuronal circuits. When a deep CNN with many layers is used for the visual system, it is not easy to compare the structure components of CNNs with possible neuroscience underpinnings due to highly complex circuits from the retina to higher visual cortex. Here we address this issue by focusing on single retinal ganglion cells with biophysical models and recording data from animals.
By training CNNs with white noise images to predict neuronal responses, we found that fine structures of the retinal receptive field can be revealed. Specifically, convolutional filters learned are resembling biological components of the retinal circuit. This suggests that a CNN learning from one single retinal cell reveals a minimal neural network carried out in this cell. Furthermore, when CNNs learned from different cells are transferred between cells, there is a diversity of transfer learning performance, which indicates that CNNs are cell-specific. Moreover, when CNNs are transferred between different types of input images, here white noise v.s. natural images, transfer learning shows a good performance, which implies that CNNs indeed capture the full computational ability of a single retinal cell for different inputs.  
Taken together, these results suggest that CNNs could be used to reveal structure components of neuronal circuits, and provide a powerful model for neural system identification.

\end{abstract}


%
\IEEEpeerreviewmaketitle

\section{Introduction}
%
%
%
%
\IEEEPARstart{D}{eep} convolutional neural networks (CNNs) have been a powerful model for numerous tasks related to system identification in recent years~\cite{Lecun2015Deep}. By training a CNN with a large set of target images, it can achieve the human-level performance for visual object recognition. However, it is still a challenge for understanding the relationship between computation and underlying network structure of components learned within CNNs~\cite{Smith_2016, Marblestone_2016}. Thus, visualizing, interpreting, and understanding CNNs are not trivial~\cite{Zeiler2014Visualizing}. 

Inspired by neuroscience studies~\cite{Hassabis2017}, a typical CNN consists of a hierarchical structure of layers~\cite{Lecun2010Convolutional}, where one of the most important properties for each convolutional (conv) layer is that one can use a conv filter as a feature detector to extract useful information from input images~\cite{Simonyan2015Very,Krizhevsky2012ImageNet}. Therefore, after learning, conv filters are meaningful. The features captured by these filters can be represented in the original natural images~\cite{Zeiler2014Visualizing}. Often, one typical feature shares some similarities with certain image patches of natural images from the training set. These similarities are obtained by using a very large set of specific images with reasonable labels. The benefit of this is that features are relatively universal for one category of objects, which is good for recognition. However, it also causes the difficulty of visualization or interpretation due to the complex nature of natural images, i.e., the complex statistical structures of natural images~\cite{Simoncelli2001Natural}. As a result, the filters learned in CNNs are often not obvious to be interpreted~\cite{Zeiler2011Adaptive}. 

On the other hand, researchers begin to adapt CNNs for studying the central questions of neuroscience~\cite{Kriegeskorte2015,Yamins2016Using}. For example, CNNs have been used to model the ventral visual pathway that has been suggested as a route for visual object recognition starting from the retina to the visual cortex and reaching the inferior temporal (IT) cortex~\cite{Yamins2013Hierarchical,Yamins_2014,Khaligh-Razavi2014,Yamins2016Using}. The prediction of neuronal responses, in this case, has a surprisingly good performance. However, the final output of this CNN model is representing dense computations conducted in many layers, which may or may not be relevant to the biological underpinnings of information processing in the brain. Understanding these network components of CNNs is difficult given that the IT cortex part is sitting at a higher level of our visual system~\cite{Yamins2016Using}.

\begin{figure*}[th]
\begin{center}
\includegraphics[width=1.9\columnwidth]{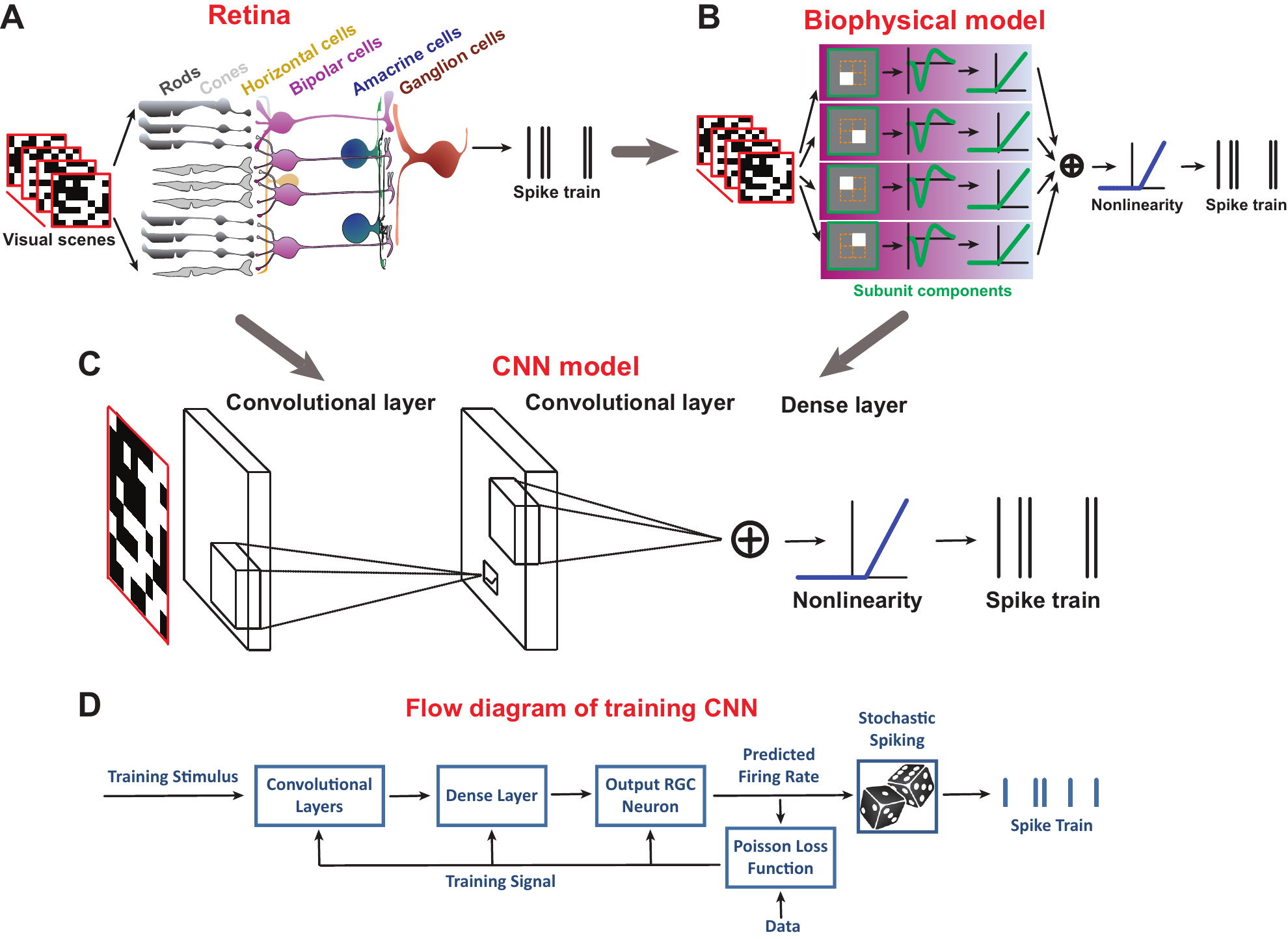}
\caption{ Illustration of biophysical model and CNN model to study the retinal computing. (A) Retinal circuit computes its output as a sequence of spikes for each RGC when visual scenes are received by the eyes. (B) Illustration of RGC model structure used in the current paper. Simplified neuronal circuit of a single RGC can be represented by a biophysical model that consists of a bank of subunit linear filters and nonlinearities. Note there are four subunits playing the role of conv filters. (C) Illustration of CNN model structure. CNN is used to train the same set of stimulus images to predict the spikes of all images for both biological RGC data and biophysical model data. (D) Flow diagram of training a CNN model to predict the RGC responses.  
}
\label{fig:retina}
\end{center}
\end{figure*}

In principle, CNN models can also be applied to early sensory systems where the organization of underlying neuronal circuitry is relatively clear and simple. Thus one expects knowledge of these neuronal circuits could provide useful and important validation for CNNs. Indeed, a few studies applied CNNs and their variations to earlier visual system, such as the retina~\cite{mcintosh2016deep, Batty2017,vance2018bioinspired, maheswaranathan2018deep}, V1~\cite{vintch2015convolutional, Antolik2016,kindel2017using,cadena2017deep,Klindt2017, whiteway2018characterizing,ukita2018characterization} and V2~\cite{rowekamp2017cross}. Most of these studies are driven by the goal that the better performance of neural response can be achieved by using either feedforward and recurrent neural networks (or both). These new approaches increase the complexity level of system identification, comparing to conventional linear/nonlinear models~\cite{McFarland2013,chichilnisky2001a,Liu_2015}. Some of these studies also try to look into the details of network components after learning to see if and how they are comparable to the biological structure of neuronal networks~\cite{maheswaranathan2018deep,Klindt2017,ukita2018characterization}.

The retina, compared to other earlier visual systems, has a relatively simple neuronal circuit with three layers of neurons as photoreceptors, bipolar cells and ganglion cells, together with inhibitory horizontal and amacrine cells in between as illustrated in Fig.~\ref{fig:retina}(A). The retinal ganglion cells (RGCs), as the only output neurons of the retina, send visual information via the optic tracts and the thalamus to cortical areas for higher cognition. Each RGC receives input from a number of excitatory bipolar cells (BCs) as driving force to generate spikes, which traditionally is modeled by a biophysical model with a number of filters and nonlinearities as in~Fig.~\ref{fig:retina}(B)~\cite{McFarland2013,Liu2017,jia2018neural}. Thus, it serves as a typical model for both deciphering the structure of neuronal circuits~\cite{helmstaedter2013connectomic,zeng2017neuronal,marc2013retinal,seung2014neuronal,sanes2015types,demb2015functional} and testing novel methods for neuronal coding~\cite{chichilnisky2001a,Pillow2008Spatio,McFarland2013}. In this study, we use the retina to show what could be learned by CNNs. Unlike these previous studies focusing on a large population of retinal ganglion cells~\cite{maheswaranathan2018deep,Klindt2017}, here we take a different viewpoint by modeling single RGC with CNN as illustrated in Fig.~\ref{fig:retina}(C). In this way, one can use the single cell to reveal the details of the receptive field of visual neurons~\cite{ukita2018characterization}.

Our aim is to study what kind of possible biological structure components in the retina can be learned by CNNs, and how one can use CNNs for understanding the computations carried out by single retinal cells.
These questions concern the research focus of understanding, visualizing and interpreting the CNNs components out of its black box.

Like those typical existing biophysical RGC models based on interpretable biophysical proprieties of the receptive filed measured experimentally \cite{chichilnisky2001a,Pillow2008Spatio}, we found our CNN model is also interpretable. By using a minimal biophysical model of RGC, we found the conv filters learned in CNNs are essentially the bipolar subunit components of RGC model. 
Furthermore, we applied CNNs to analyze biological RGC data recorded in the salamander. The conv filters are resembling the receptive fields of bipolar cells that sit in the previous layer of RGC and pool their computations to a downstream single RGC. 
Such a fine structure picture of the retinal receptive field revealed by CNNs suggests that convolutional filters learned are resembling biological components of the retinal circuit. Thus a CNN learning from one single retinal cell reveals a minimal neural network carried out in this cell. 

In addition, different RGCs were trained to get their corresponding CNN models, then we transferred these CNNs between these RGCs to test the ability of transfer learning.  There is a diversity of transfer learning performance. In general, the CNN models have a good performance in transfer learning. However, the best performance is still the one trained with its own original data, which implies that the CNN model is cell-specific with a set of filters inherited from the target RGC. Furthermore, when CNNs are applied to a different input domain, here white noise images v.s. natural images, they can capture the meaningful responses of the RGCs in terms of the proper number of spikes and right spike timings of spikes, even in the cases where there is no spike for some specific images. This implies that CNNs indeed captures the full computational ability of one cell for different inputs.  

Some preliminary results of this study were presented in a NIPS workshop short communication~\cite{yan2017revealing}.

\section{Methods}

\subsection{Biophysical RGC model}
A biophysical RGC model as in Fig.~\ref{fig:retina} (B) was modeled as a typical subunit model used previously~\cite{gollisch2008rapid, McFarland2013}. The model cell has four subunits with a spatial filter of the size 2x2 pixels, similar to a conv filter of CNNs but only sitting at a specific spatial location, and a temporal filter to take into account of temporal dynamics. Each subunit convolves the incoming stimulus image and then applies a nonlinearity of threshold-linear rectification. The output subunit signals are then polled together by the RGC. The polled signal is applied with a threshold-linear output nonlinearity with a positive threshold at unity to make spiking sparse. Thus, with this model, a given sequence of stimulus consisted of white noise images with the size as 8x8 pixels can generate a train of spikes.

\subsection{Biological RGC data}
A public dataset of RGCs recorded in salamander as described in~\cite{Onken_2016, vance2018bioinspired, Liu2017} was used for CNNs modeling. Briefly, a population of RGC spiking activities was obtained by multielectrode array recordings as in~\cite{Liu_2015}. 
The retinas were optically stimulated with spatiotemporal white noise images, temporally updated at a rate of 30 Hz and spatially arranged in a checkerboard layout with stimulus pixels of 30x30 $\mu m$. The recording time is about 4 hours so that there are enough stimulus images and spikes for training a CNN model. A dataset of 300 natural images was also used as the stimulus for studying natural image responses~\cite{Onken_2016,Liu2017}. 

\subsection{CNN RGC model}
We used naive CNN models containing a different number of convolution layers and a single dense layer as illustrated in Fig.~\ref{fig:retina} (C). CNNs were implemented with Keras using Tensorflow as backend. The code is available at https://sites.google.com/site/jiankliu.

To model biophysical RGC model data, the conv filter size was fixed as $7\times7$. Both the number of filters and layers were tested. To model biological RGC data, several sets of parameters in convolution layers, including the number of layers, the number, and size of convolution filters were explored. The prediction performance is robust against these changes of parameters. Therefore we adopted the filter size of $15\times15$ in the first conv layer and $3\times3$ in the second conv layer.

For training CNNs with biophysical RGC model data, we generated a data set consisting of 600k training samples of white noise images, and an additional set of 10k samples for testing. The training labels are a train of binary spikes with 0 and 1 generated by the model. 

For biological RGC data recorded in the salamander, there are about 320k training samples of white images and labels as the number of spikes as in [0 5] for each image. The test data have 300 samples, which were repeatedly presented to the retina for about 200 trials.

The procedure of training a CNN RGC model is shown in Fig.~\ref{fig:retina} (D), where either binary spike train or firing rate of biophysical and biological test data can be compared to the predicated firing as CNN output for calculation of the Pearson correlation coefficient (CC) as a performance measure. A Poisson loss is used to optimize the CNN output to match the spiking labels. The final nonlinearity after the dense layer is a standard soft-plus function. 

For preprocessing of data, a standard technique of spike-triggered average was applied to get a 3D spatiotemporal receptive field filter~\cite{chichilnisky2001a}. The singular value decomposition of this 3D filter yields a temporal filter and a spatial receptive field~\cite{Gauthier_2009}.

Two versions of CNNs were used. (Version I) The first version of CNN has only spatial filters without temporal filters to be fitted. For this, data was temporally correlated first by convolving every pixel of the whole set of stimulus images with the temporal filter first along the temporal dimension~\cite{Kaardal2013,Liu2017}. In this way, a sequence of spatial images was obtained as inputs with the corresponding spike train as output labels for CNN model, such that this CNN makes analysis focusing on the spatial structure of receptive fields. In addition, this CNN has much fewer parameters, for example, when the temporal filter of interest is lasting for 600 ms with 30 Hz, then CNN parameters are 20 times less. (Version II) The second version of CNN is the full model with both spatial and temporal filters to be learned from the RGC data. This usually results in a large number of parameter that causes unavoidable problems for traditional statical models~\cite{Pillow2008Spatio, McFarland2013}. The recent advancements of deep learning make it relatively easier to fit the data with high-dimensional parameters. Both versions of CNNs were used and compared to see the effects on the performance. 

\begin{figure}[th]
\includegraphics[width=\columnwidth]{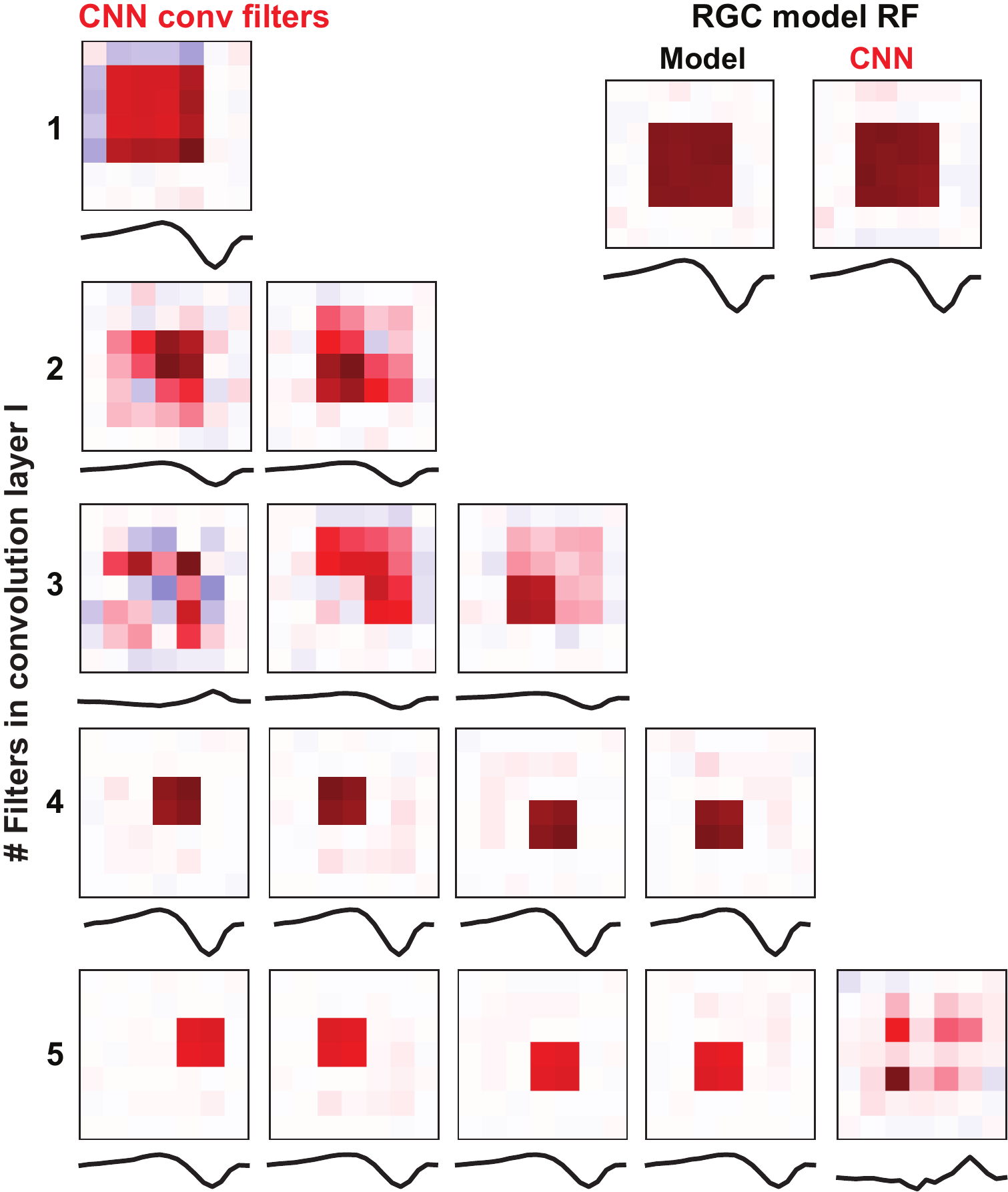}
\caption{Subunit structure of RGCs model data revealed by CNNs. Visualizing conv filters learned in a CNN with one layer of convolutional filters. The number of conv filters is from 1 to 5, where both spatial and temporal filters are learned by CNNs. (Inset) spatial receptive field and temporal filter of the modeled RGC computed by spike-trigger average (left) and CNN (right). Note the filter size is $7\times7$, and the full size of the receptive field is $8\times8$. }
\label{fig:02}
\end{figure}

When CNNs are used for object recognition task by learning a set of natural images, it is important to visualize what kind of features of natural images are learned by conv filter~\cite{Zeiler2014Visualizing}. Similarly, here one can also visualize the image features represented by each CNN conv filter. The feature here means the response-weighted average feature, which is the spike-triggered average generated by a batch of random noise stimulus and its average activation from the corresponding feature map in the first layer.

\begin{figure}[bth]
\includegraphics[width=1.\columnwidth]{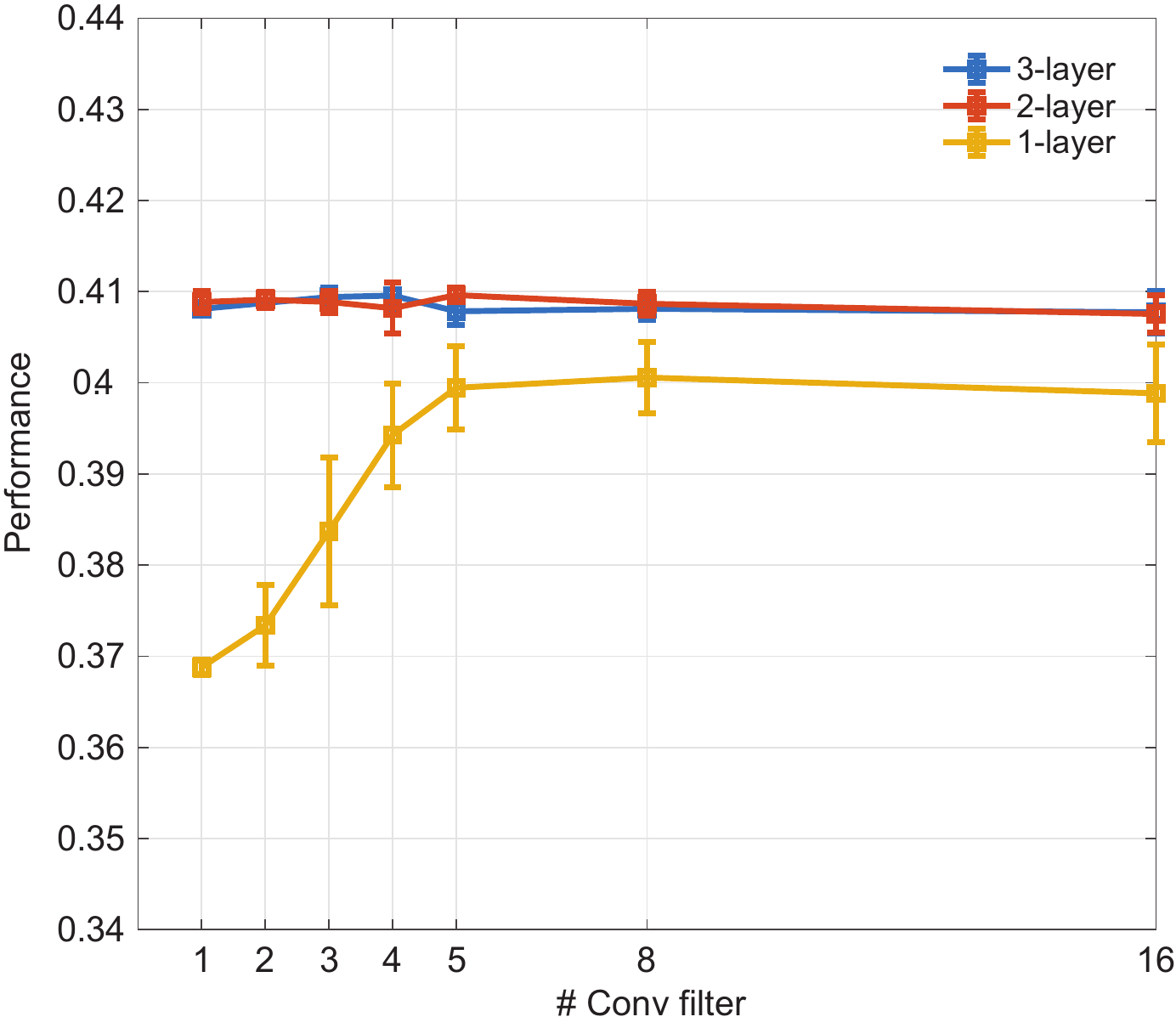}
 \caption{
  CNN performance saturated when there are more conv filters and layers. Data points represented by 10 runs with mean $\pm$ standard derivation (STD).
 }
\label{fig:layer}
\end{figure}

\begin{figure*}[th]
\includegraphics[width=1.9\columnwidth]{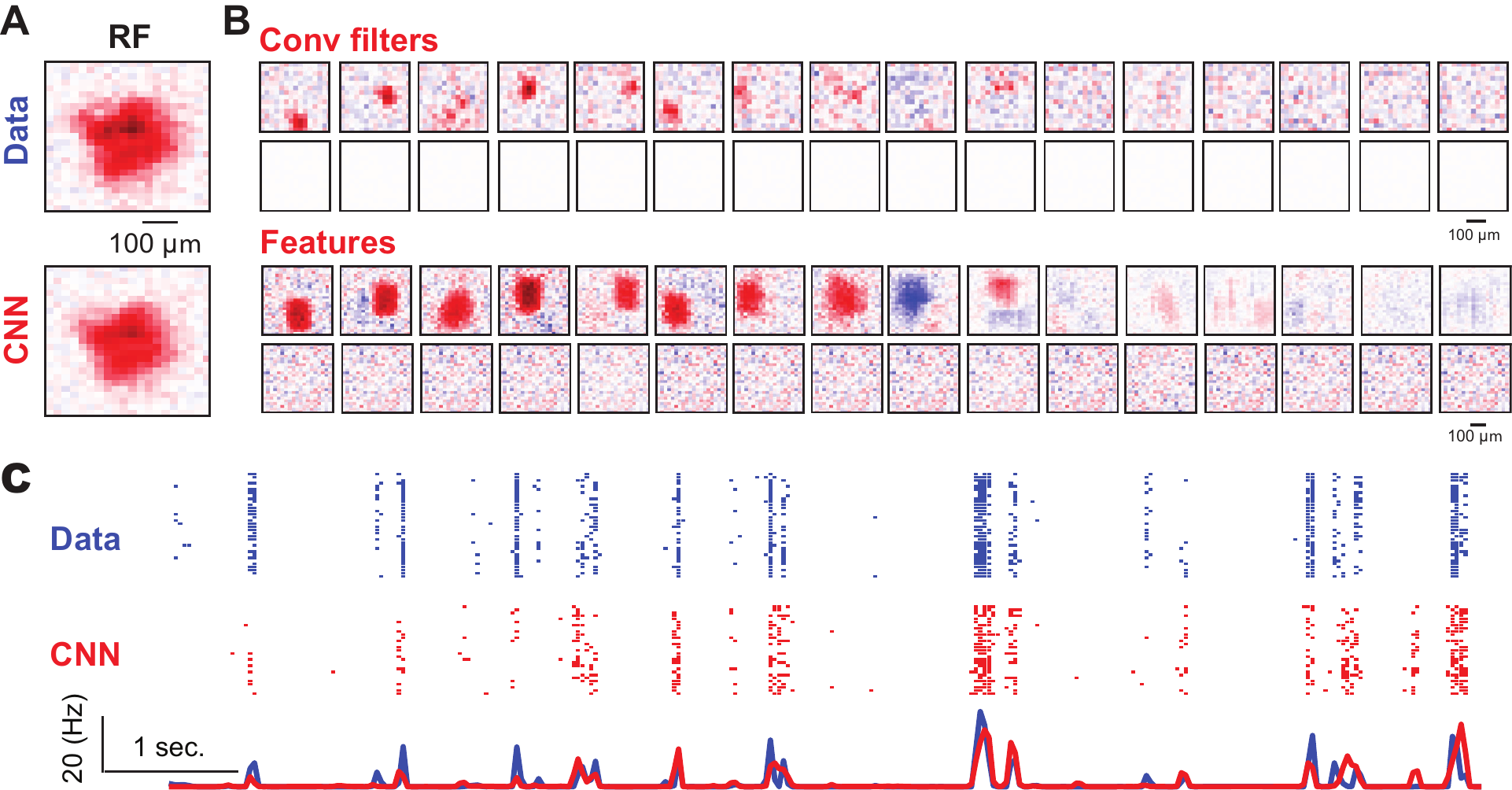}
\caption{ Subunit structures of biological RGC revealed by CNN. (A) Receptive fields of one RGC computed by STA and CNN prediction. (B) Visualizing CNN model components of conv filters and average features represented by each filter. (C) Neuronal response predicted by CNN visualized by RGC data spike rasters (upper), CNN spike rasters (middle), and their average firing rates (bottom).}
\label{fig:cell_filter}
\end{figure*}

\begin{figure}[tbh]
\begin{center}
\includegraphics[width=1\columnwidth]{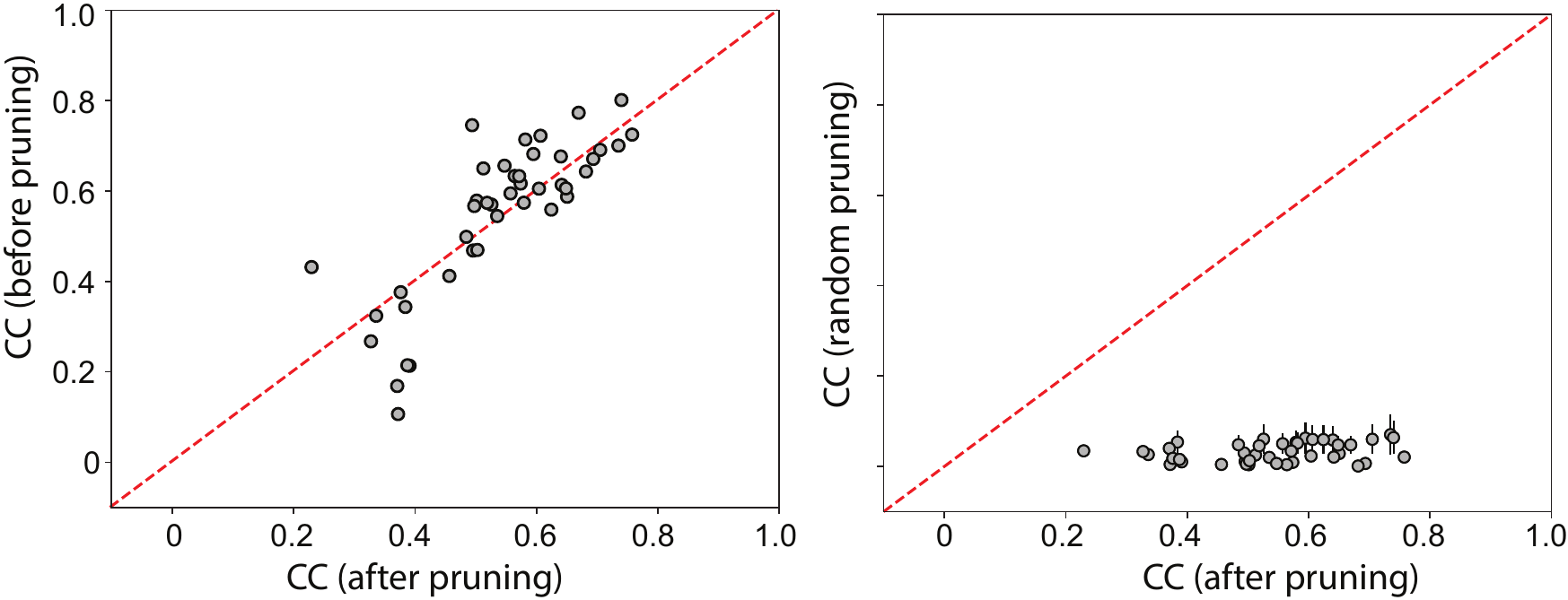}
\caption{Pruning convolutional filters in CNNs. (Left) CNN model performance (CC) maintained at a similar level after pruning CNN by using only a subset of effective (non-zero) filters as in Fig.~\ref{fig:cell_filter}. (Right) CNN performance dropped to zero when pruning CNN with the same number of parameters but a randomly selected subset of filters. Error bars indicate 10 random prunings (mean$\pm$STD). }
\label{fig:prune}
\end{center}
\end{figure}

\section{Results}

By using both clearly defined biophysical model and real retinal data, we show that CNNs are interpretable when single RGC was modeled with the benefit to clarify what has been learned in the network structure components of CNNs. 
Recently, a variation of non-negative matrix factorization was used to analyze the RGC responses to white noise images and identify a number of subunits as bipolar cells of one RGC~\cite{Liu2017}. With this picture in mind, here we address the question of what types of network structure components can be revealed by CNN when it is used to model the single RGC response.

\subsection{Subunits of modeled RGC as CNN filters}

We set up a biophysical RGC model with four subunits as in Fig.~\ref{fig:retina}(B), which is resembling a 2-layer network with one layer of subunits and one layer of single RGC. By using a set of white noise images, the model generated a sequence of spikes to simulate a minimal neural network of the retinal ganglion cells. With the input of stimulus images and the output of RGC spikes, we can train a CNN as in Fig.~\ref{fig:retina}(C) to predict the simulated spikes generated by the RGC model. 

Given that our focus is looking into the structure of network components learned in CNNs, we varied a number of parameters to train the CNNs, in particular, the number of conv filters from 1 to 16. When only one conv filter is used, the learned filter has a similar structure as the receptive field of the modeled RGC as in Fig.~\ref{fig:02} that can be obtained by the standard method termed spike-triggered average~\cite{chichilnisky2001a} (see Methods). When there are more conv filters, there are more fine structure of the receptive field as filters learned by CNNs. We found that when training the CNN with four conv filters, the outcome filters resemble the subunits used in the biophysical RGC model. The outcome filters learned by the CNNs are convergent in the sense that there are only four ``effective" filters similar to the model subunits, the rest of the filters are noise, for example when there are five conv filters in Fig.~\ref{fig:02}. This observation is similar to a recent study where non-negative matrix factorization was used for identifying the subunits~\cite{Liu2017}.

The nature of the black box of CNNs often forces researchers to tune the parameters of CNNs according to the performance that is usually the accuracy of the tasks, such as image classification. Here, the performance is then the correlation between CNNs output and modeled RGCs spiking response. Not surprisingly, CNNs can give a good performance for predicting the RGCs response, which is consistent with the previous studies of various types of visual neurons~\cite{mcintosh2016deep,Yamins2016Using,Klindt2017}.
More importantly, here we also found when there is enough number of conv filters, increasing the number of conv filters does not make the performance better as shown by an evolution of filter change with an increasing number of filters as in Fig.~\ref{fig:layer}. The performance is convergent when the number of conv filter reaches 5. 

In addition, the number of conv layer is tuning the CNN performance to reach the saturation level. There is no difference when there are two layers or three layers of CNN conv filters. When there are 2 or more layers of conv filters, the performance is stabilized independently of the number of conv filters used. In either case, when there is only one conv filter used, it yields a result where the receptive filed of RGC is learned as one conv filter of CNN.  

Therefore, these results show that increasing the number of conv filters and layers does not increase the performance when enough components are used to capture the underlying biophysical properties of RGC model. In other words, CNN parameters could be highly redundant when setting up to a larger number. Such a redundancy of parameters is widely observed for deep learning models~\cite{Denil_2013,Han2015Deep}. This point will be shown by the biological data below in details.


Altogether, These results suggest that CNNs can identify the underlying hidden network structure components within the RGCs model by only looking at the input stimulus images and the output spiking response.

\begin{figure}[th]
\includegraphics[width=1\columnwidth]{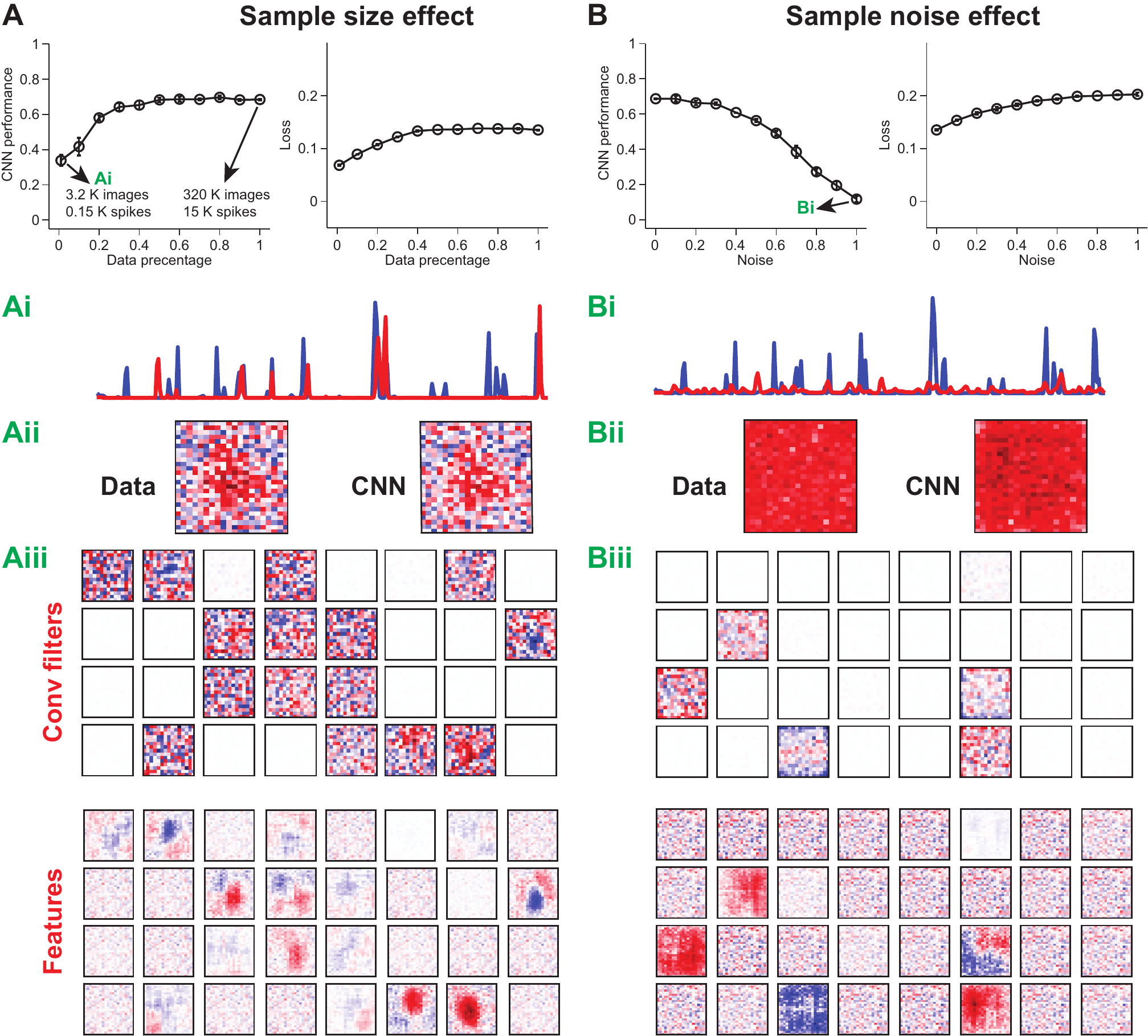}
\caption{Effects of training sample size and noise on the performance of CNNs. (A) Effect of sample size changing from 1\% (0.01) to 100\% (1) of the whole training set images on the performance of CNN (CC, upper) and the loss of training (Loss, bottom), where 1\% of training data has about 3.2K images and 0.15K spikes (3.05K of non-spiking labels as zero) due to the spare firing property of RGC. Data point of Ai is shown in (Ai-Aiii). (Ai) CNN output v.s. test data firing rate. (Aii) receptive field of data (left) v.s. CNN prediction (right). (Aiii) conv. filters and features leaned by CNN. (B) Similar to (A) but for the effect of noise in sample images, where the ratio of noise is changing from 0\% (without noise) to 1 (100\% of noise without data).  (Bi-Biii) Similar to (Ai-Aiii).} 
\label{fig:021}
\end{figure}

\subsection{Subunits of biological RGC as CNN filters}

To further characterizing the structure components of CNNs in details, we use CNNs to learn the biological RGC data with similar images of white noise and spiking responses. We first use CNNs to study temporally correlated RGCs data where no temporal filter is needed to be learned by CNNs (see Version I CNN model in Methods) with the benefit of fewer parameters of CNNs to be learned. Similar to the results of the RGC model above, the outputs of CNN model can recover fine structures of the receptive field of RGC data very well as in Fig.~\ref{fig:cell_filter}(A). We also found that the learned conv filters converge to a set of localized subunits whereas the rest of filters are noisy and close to zero as in Fig.~\ref{fig:cell_filter}(B). The size of these localized filters is comparable to that in bipolar cells around 100 $\mu m$ \cite{Liu2017}. 

In addition, the features (see Methods) represented by these localized conv filers are also localized. Given the example RGC is an OFF type cell that responds to the dark part of images strongly, most features have similar OFF peaks resulted from the OFF BC-like filters. These OFF features tile the space of the receptive field of RGC. Interestingly, there are some features with ON peaks, which play a role as inhibition in the retinal circuit. A few features have some complex structures mixed with OFF and ON peaks, which are mostly resulted from the less localized filters. However, if the filters are pure noise, the resulting features are pure noise without any structure embedded. Besides filters and features, the CNN model generates a good prediction of RGC response as in Fig.~\ref{fig:cell_filter}(C). These observations are similar across different RGCs recorded. 

When there are 32 conv filters used in the CNN, there are enough number of conv filters to fit the data. Given there are many redundant filters (minimal 16 filters close to zero in Fig.~\ref{fig:cell_filter} (B)) unused in CNN learning, they unlikely reflect any interesting biophysical properties of RGC data, so could be no contribution for the performance of CNN. Indeed, CNN performance maintains at a similar level when the conv filters are pruned such that the only effective filters are used in CNN. The selection of effective filters was quantified with the spatial autocorrelation of conv filters~\cite{Liu2017}. The pruning results of a population of RGCs show that the performance is similar to a small subset of effective conv filters in Fig.~\ref{fig:prune}. When pruning is done with randomly selected conv filters out of 32, the performance drops to zero as in Fig.~\ref{fig:prune} (right).

These results confirm the observation shown in the biophysical model RGC, where the CNN performance cannot be increased with more parameters than enough. In terms of biological RGCs, the subunits are the upstream bipolar cells~\cite{Liu2017}. Thus, the conv filters play a functional role as bipolar cells when using CNNs to model biological RGCs.

Next, we examine the effect of sample size and noise on CNN model. For the same cell shown in Fig.~\ref{fig:cell_filter}, the size of samples/images was changed in a wide range from 1\% to 100\%, in this way, the corresponding number of spikes recorded in RGC is reduced to about 150 spikes with 1\% of training data. Both CNN performance and loss after training are dependent on how much data used for training as in Fig.~\ref{fig:021}(A). Surprisingly, with only about 150 spikes from 1\% of data, we can still obtain some level of performance with CC about 0.35 as in Fig.~\ref{fig:021}(Ai). Although the receptive field with such small amount of spikes is not good (Fig.~\ref{fig:021}(Aii)), the performance of CNNs do not drop that much (CC is 0.35 v.s. 0.75, but with 150 v.s. 15K spikes). Thus, CNNs seems to need only a relatively small set of spikes for training to get a reasonable performance. Note with 30\% of data (4.5K spikes), the performance is almost similar to the full data. In this sense, the CNNs seem to be much less data-demanding than traditional biophysical RGC models~\cite{McFarland2013,Pillow2008Spatio}. The resulting filters and features of CNNs are also worse, however, the spareness of filters still holds although 32 filters are used.

In contrast to the sample size, the noise has a much larger effect on the CNN model as in Fig.~\ref{fig:021}(B). Keeping the sample size unchanged, we replaced part of data images with irrelevant noise images, for instance, 70\% of noise means there are 30\% of data and 70\% of noise images. Although the data percentage is the same as 30\%, the CNN trained with noise has a much worse performance (CC $\approx$ 0.4) comparing to the same amount of data without noise (CC $\approx$ 0.7).  Fig.~\ref{fig:021}(Bi-Biii) shows an example result of training CNN with complete noise images. Even in this case, the redundancy of filters makes most filters close to zeros.

\subsection{Transfer learning across different RGCs}

Given there is a population of RGCs recorded experimentally and then modeled by CNNs, one can study the behavior of transfer learning, or the generalization ability of CNNs model, i.e., using the CNN model learned from one RGC to predicate the response of another RGC. Several scenarios of transfer learning are commonly used in deep learning, including using a pre-trained model directly, fine-tuning a pre-trained model, or fixing features of a pre-trained model but adjusting dense layer~\cite{Yosinski2014}. Here we took the approach to use a pre-trained CNN directly from one RGC to other different RGCs. This is suitable for our experimental setup, as a large full-size of white noise images were presented to all RGCs of a population at one time. Different RGCs are sitting at different spatial locations of images, therefore they are seeing parts of the whole image. However, due to the nature of white noise images, the statistics of the ensemble input images are the same, or at least closely similar to Gaussian, across different RGCs. Thus, one expects that the transfer learning of CNN model, in this case, has a good performance.

\begin{figure}[tht]
\includegraphics[width=\columnwidth]{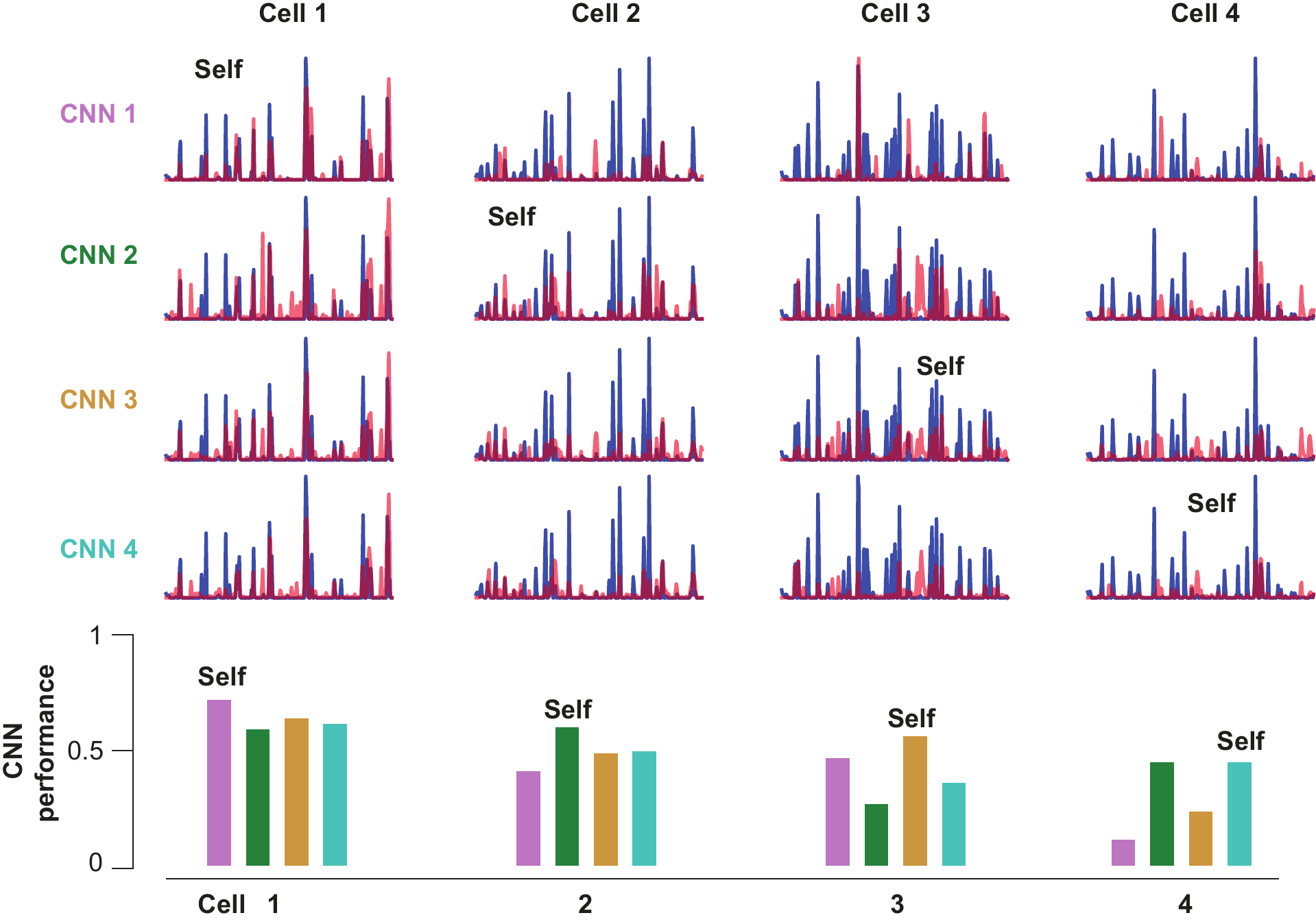}
\caption{ CNNs transfer learning across different cells. (Upper) Four example cells used to train the CNN models. Learned CNNs are then transferred across different RGCs for prediction. Over this 4x4 matrix, the diagonal ones are the cases of self-training (Target to Target) with each cell's data. Cell 1 is the same cell as in Fig.~\ref{fig:cell_filter}. CNN 1 is the model obtained by cell 1. The first row represents that CNN 1 model is transferred to predict the test data of Cell 2-4. The first column represents that three CNN models (CNN 2-4) from the other three cells (Cell 2-4) are transferred to predict the test data of Cell 1. (Bottom) Performance of this 4x4 matrix. The first four points are calculated for cell 1 from CNN 1-4. Different CNNs are colored in different colors. }
\label{fig:03}
\end{figure}

However, we found there is a large diversity of transfer learning performance across different RGCs as shown in Fig.~\ref{fig:03}, where there are four example cells showing their CNNs model predictions (diagonal traces labeled as ``Self") and the corresponding transfer learning behaviors (off-diagonal traces). The CNN performance characterized by CC shows the tendency that the best performance is always from the CNN model trained with their own cell. For instance, cell 1 has a good test performance with its own CNN model, denoted as CNN 1. When CNN 1 is transferred to the other three cells, the prediction power is lower than cell 1. However, the CNN 3 is transferred to cell 1 with a better performance of CC=0.64 comparing to CC=0.56 for cell 3 itself. Interestingly, all CNN models have a reasonably good performance for cell 1, even though that CNN 4 has a fairly good performance for cell 4 itself. 

\begin{figure}[tht]
\includegraphics[width=\columnwidth]{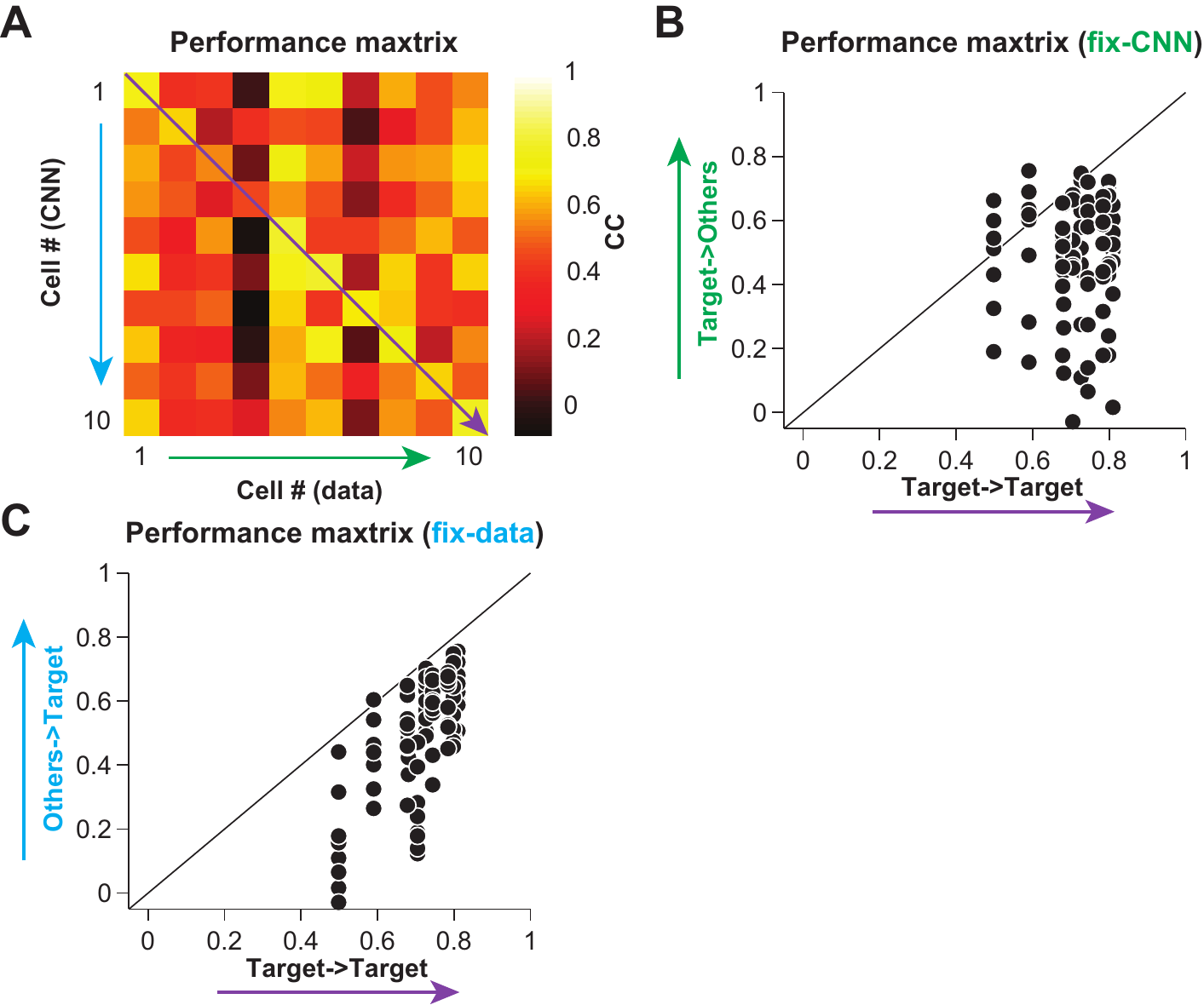}
\caption{CNNs transfer prediction for a population of 10 RGCs. (A) Performance matrix of 10 CNNs across different RGCs, where the diagonal ones are self-training (Target to Target, purple), each row is transferring the CNN trained by target cell to all other cells (Target to Others, fix-CNN, green), and each column is transferring all other CNNs to the target cell (Others to Target, fix-data, light blue). The first four cells are the same ones as in Fig.~\ref{fig:03}. The rows and columns of performance matrix have the same meaning as in Fig.~\ref{fig:03}. (B) Performance matrix shown as a scatter plot of self-training v.s. fix-CNN. (C) Performance matrix shown as a scatter plot of self-training v.s. fix-data.  }
\label{fig:04}
\end{figure}

To further investigate the ability of transfer learning of CNN model, we collect a population of 10 RGCs (including those four cells in Fig.~\ref{fig:03}). The population plots in Fig.~\ref{fig:04} confirm the observation above. The scatter plot of self-learning (Target $ \rightarrow $ Target) v.s. transfer-learning (Target $ \rightarrow $ Others) in Fig.~\ref{fig:04} (left) shows that the CNN model has a reasonable good performance for both self-learning and transfer-learning, yet the results are quite diverse. Note that even for the worst cell with lowest CC in self-learning, when its CNN is transferred to other cells, its CNN has a better performance. 

Similarly, as above, the performances are always lower when other CNNs are transferred to the target cells themselves (Fig.~\ref{fig:04} (right)). This indicates that each CNN model is really optimized for the target cell after training. For the best cell who has highest CC by CNN, other CNNs trained with other cells also have good performance in general. In contrast, for the worst cell who has lowest CC by CNN, other CNNs trained with other cells also have bad performance. 

These results suggest that the CNN learned from each cell is very specific to that particular cell. The resulting CNNs, therefore, learns to obtain a minimal neural network that carries out the essential computations done by that cell.


\begin{figure}[tht]
\centerline{\includegraphics[width=1\columnwidth]{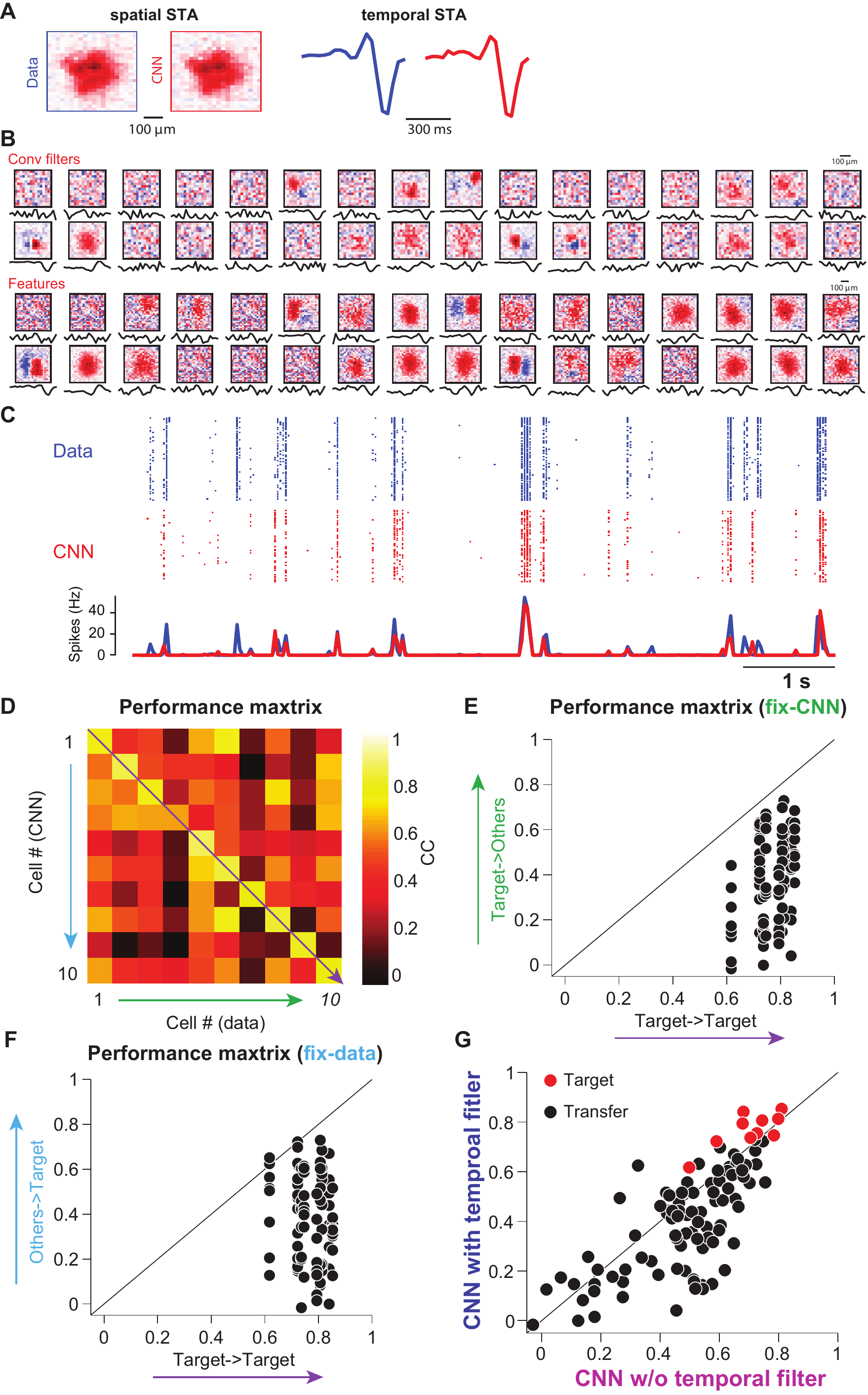}}
\caption{ Spatial and temporal filters of biological RGC data revealed by full CNN model. (A) Receptive fields as spatial STA of the example cell and CNN prediction. Temporal filters are also recovered by CNN. (B) Visualizing CNN model components of both conv filters and average features represented by each filter in both spatial and temporal dimension. (C) Neuronal response predicted CNN visualized by RGC data spike rasters (upper), CNN spike rasters (middle), and their firing rates.
 (D-F) Transfer prediction by full CNN models for a population of 10 RGCs. All plots have the same meanings as Fig.~\ref{fig:04}. (G) CNN prediction improved with temporal filter included, but transfer prediction is worse in general. Performance (CC) matrix shown as a scatter plot of CNN without temporal filter (Fig.~\ref{fig:04}) v.s. full CNN with temporal filter. Black indicates transfer prediction, red indicates target prediction.  }
\label{fig:02time}
\end{figure}

\begin{figure*}[th]
\centerline{\includegraphics[width=1.8\columnwidth]{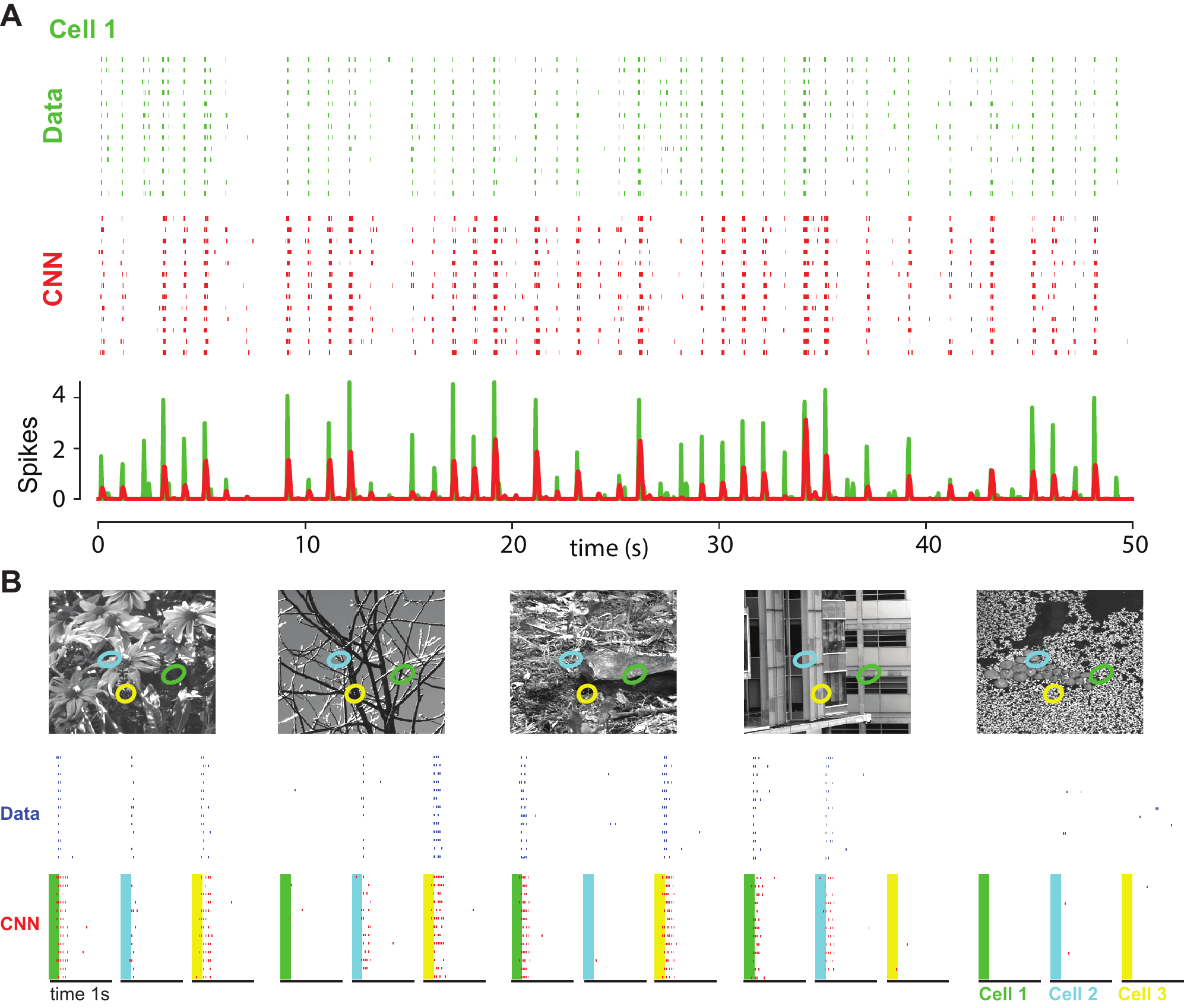}}
\caption{Transfer prediction of natural images by CNNs trained with white noise images. (A) Performance of CNN for the example RGC (green cell in (B)) with 50 images (one image per second). (B) Performance of CNN with four example images and three RGCs. (top ) five images overlaid with the outlines of the receptive field of three RGCs colored in green, light blue and yellow. (bottom) spike response of each GC for each image (blue) together with spikes sampled from CNN output (red). Each image was presented for 200 ms long (colored shadow window) then followed by 800-ms gray period. Note RGC response is delayed after the onset of an image. Each image triggers one RGC in a different manner with spiking or non-spiking depending on the texture of the image and specific RGC. }
\label{fig:image}
\end{figure*}

\subsection{Full CNN model for biological RGC data}

The results above on biological RGC data were studied by a CNN without temporal filter learned. Now we consider the full CNN model (Version II CNN model, see Methods), where both temporal and spatial filters are needed. Similar to the RGC model data, CNN can recover both spatial filters as receptive field and temporal filters as shown in Fig.~\ref{fig:02time} (A). With the full CNN model, there are 20 times more parameters than that of temporally correlated data. Both conv filters and features visualized in CNN have still a located spatial structure and good temporal filer shape for a subset of filters. Not surprisingly, the prediction of CNNs for neuronal response has a good performance as well.

Similar to Fig.~\ref{fig:04}, we also tested the transferring learning ability of the full CNNs model for the same population of 10 RGCs. The results in Fig.~\ref{fig:02time} show a similar tendency of transferring learning for the CNN model. Yet, there are some differences between the two versions of CNN models, which can be seen by comparison in Fig.~\ref{fig:02time}. For the target case, where each CNN was trained by using that particular RGC data, the full CNN yields a better performance than the CNN without temporal filter learned. However, when full CNNs are transferred between different cells, their performance is in general worse than reduced CNNs. That indicates that the full CNN with both spatial and temporal filters is more specific to the particular cell used for training. In turn, such a cell-specific CNN can not be used to explain other cells. Therefore, this confirms the result that CNN indeed learns the underlying neuronal computation carried out by the biological cell, which can not be transferred across different cells. 

Given there are 20 times more parameters in the full CNN, the conv filters have less clearly localized structures than those in the CNN of reduced temporal correlations when comparing the filters in Fig.~\ref{fig:cell_filter} and Fig.~\ref{fig:02time}. This seems to be caused by the limited sample size of biological RGC data. When the biophysical RGC model is used for both versions of CNN, there is no difference in terms of the structure of conv filters, see the results shown in~\cite{yan2017revealing}.  
Therefore, depending on the questions to be addressed, one may want to choose the simple or full version of CNN for modeling static images or dynamical videos, as videos contain strong temporal correlations that have a strong impact on the adaptation of neuronal dynamics~\cite{Liu_2015}.

\subsection{Transfer learning between different stimulus images}

Above we tested transfer learning of CNNs across different cells. Here we further test transfer learning of CNNs from each cell but for different input images. So far, the CNN model of each GC trained by white noise images, one can test the ability of transfer prediction by directly applying learned CNNs to natural scenes. 

For this, a sequence of 300 natural images was presented to the retina, where each image was briefly presented for 200 ms and followed by 800 ms empty scene. Such a protocol of stimulation is to leaving out temporal adaption for each image (see details in \cite{Onken_2016}). Therefore, we used a temporal decorrelated CNN model (Version I) to fit the neuronal response of this sequence of flashing images. 

Surprisingly, CNNs learned by white noise images can predict the spiking response of natural images quite well. Fig.~\ref{fig:image}(A) shows the result of one example cell. For clear illustration, we only show a partial sequence of 50 images. The detailed result is demonstrated with five images together with three cell in Fig.~\ref{fig:image}~(B). Each RGC reads a different part of images as the receptive field is located differently. Interestingly, CNNs predict the responses well when all three cells fire for image 1, in addition, CNNs can also predict when one of three cell is not firing with spikes for image 2-4. Even when all three cells are not firing completely for image 5, the outputs of CNNs are also silent. 

These resulting of transfer learning across image domains, compared to those across different RGCs, suggest that CNNs intend to learn a cell-specific neural network in which convolutional filters play a role as upstream subunit cells that connect to a particular RGC. Thus, CNNs serve model of neural system identification to reveal the underlying computations of the retinal ganglion cells.  

\section{Summary \& Discussions}

In recent years, system identification of neural coding based on neural networks has been greatly improved by the development of deep convolutional neural network~\cite{Lecun2015Deep,Kriegeskorte2015, Yamins2016Using}, where multiple layers and huge banks of local subunit filters within the network are significant characteristics. Besides its powerful ability to many practical tasks~\cite{Lecun2015Deep}, the underlying structure is mimicking a hierarchical organization of the brain~\cite{Kriegeskorte2015,Yamins2016Using}. However, there is no first principle about designing the structure of a hierarchical deep learning network~\cite{Smith_2016}. Recent works begin to look into the details of network structure, in particular, with the potential connections to the biological brain~\cite{Marblestone_2016,Yamins2016Using}.

Hereby focusing on single retinal ganglion cell, we found that CNNs can learn their parameters in an interpretable fashion, and CNNs network components are close to the biological underpinnings of the retinal circuit. With the benefit of the relative well-understood retinal circuit, our results suggest that the building-blocks of CNNs are meaningful when they are applied to neuroscience for revealing network structure components. 

Deep CNNs are useful for modeling the abstract level of vision information~\cite{Kriegeskorte2015,Yamins2016Using} and neural coding in general~\cite{glaser2017machine} in neuroscience. Our current study simplified the approach used by the previous studies \cite{Kriegeskorte2015,Yamins2016Using,Yamins_2014} for the higher part of the visual cortex, where interpretable structure components of deep CNNs are difficult due to the complex hierarchical layers from the retina to the IT cortex in the brain. By training CNNs with white noise images, our current work also simplified the interpretation, comparing to the studies where natural images were used for model visual neurons with CNNs~\cite{mcintosh2016deep,kindel2017using,ukita2018characterization}, since white noise images have the benefit to mapping out the receptive field of the visual neurons~\cite{chichilnisky2001a,Liu2017}.

Unlike most of the recent studies, which are driven by the goal that the better performance of neural response can be archived by using either feedforward and recurrent neural networks (or both), here we focus on the fine structure of the receptive field in the retinal circuit. Along with several recent papers~\cite{Klindt2017,ukita2018characterization, ecker2018rotation}, characterizing the receptive field of visual neurons is important for understanding the filters leaned by the CNNs. Given the retina has a relatively clear and simple circuit, and the eye has (almost) no feedback connections from the cortical cortex, it is a suitable model system as a feedforward neural network, similar to the principle of CNNs. Certainly, the contributions from the inhibitory neurons, such as horizontal cells and amacrine cells, play a role for the function of the retina. In this sense, the potential neural networks with lateral inhibition and/or recurrent units are desirable~\cite{mcintosh2016deep,Batty2017}. 

Our approach is suitable to address other difficult issues of deep learning, such as transfer learning, since the domain of images seen by single RGCs is local and less complicated than the global structure of entire natural images. In the first case, it is surprising that transfer learning across different RGCs is not perfect given that the stimulus distribution is the same since the ensembles of white noise image stimuli for all RGCs converge to a naive Gaussian distribution. However, a further thought indicates that the CNNs learned from each cell is rather specific with a particular set of filters in convolutional layers. In the current work, we only explore the filters of CNNs. Future work is needed to investigate other components, such as nonlinearity, of CNNs.  

Here we also test the ability to transfer learning across different types of stimulus, i.e., transfer between white noise images to natural images. Traditionally, dynamical videos have been used in the studies of using CNNs with the neuronal signals of interest as fMRI data~\cite{wen2017neural}, for neuronal spiking response, a sequence of flash images is often used~\cite{Yamins2016Using, mcintosh2016deep, Antolik2016}. One needs to consider the case how to use CNNs for neuronal spikes to study dynamical visual scenes, i.e., continuous videos~\cite{Zhang2020}. Future work is needed in this direction of studying the retinal computation under dynamical natural scenes.

CNNs here modeled for the retinal RGCs can be further used for reconstruction of natural images based on the responses of retinal RGCs~\cite{Zhang2020,Parthasarathy2017}. Embedding the encoder/decoder into the retinal prosthesis has been suggested as a promising direction for visual restoration~\cite{Nirenberg2012}. Such an approach of studying spike encoding and decoding of visual scenes with neural spikes will be crucial for the next generation of neuromorphic computing, including artificial visual system~\cite{Yu2020}, where the data format processed on chips are digital spikes~\cite{dong2017spike}. One expects that the close interaction of the algorithms based on spike data and CNNs~\cite{hu2018spiking,xu2018csnn,xiao2018spike, Xu2020} with neuromorphic chips~\cite{esser2016convolutional,davies2018loihi} will greatly expand our computing capacity. 

\section*{Acknowledgment}
We thank other members of the National Engineering Laboratory for Video Technology at Peking University for helpful discussions.

\ifCLASSOPTIONcaptionsoff
  \newpage
\fi

\bibliographystyle{IEEEtran}
\bibliography{ref_all_liu}
%

%




\end{document}